\documentclass[12pt]{article}

\usepackage[totalheight = 23cm, totalwidth = 17cm]{geometry}
\usepackage{amssymb,amsmath,amsfonts,amsbsy,graphicx}

\def\mathbi#1{\textbf{\em #1}}

\newcommand{\mpl}{m_{\rm Pl}}
\newcommand{\calO}{{\cal O}}
\newcommand{\calP}{{\cal P}}
\newcommand{\calR}{{\cal R}}

\begin{document}

\begin{titlepage}

\begin{center}

\hfill APCTP Pre2017-001\\

\vskip .75in

{\LARGE \bf Correlated primordial spectra \\ in effective theory of inflation}

\vskip .75in

{\large Jinn-Ouk Gong$^{(a,b)}$ and Masahide Yamaguchi$^{(c)}$}

\vskip 0.25in

\textit{
   ${}^{a}$ Asia Pacific Center for Theoretical Physics, Pohang 37673, Korea
   \\
   ${}^{b}$ Department of Physics, POSTECH, Pohang 37673, Korea
   \\
   ${}^{c}$ Department of Physics, Tokyo Institute of Technology, Tokyo 152-8551, Japan
}

\end{center}
\vskip .5in

\begin{abstract}

We derive a direct correlation between the power spectrum and bispectrum
of the primordial curvature perturbation in terms of the Goldstone mode
based on the effective field theory approach to inflation. We show
examples of correlated bispectra for the parametrized feature models
presented by the Planck collaboration. We also discuss the consistency
relation and the validity of our explicit correlation between the power
spectrum and bispectrum.

\end{abstract}

\end{titlepage}

\renewcommand{\thepage}{\arabic{page}}
\setcounter{page}{1}
\renewcommand{\thefootnote}{\#\arabic{footnote}}
\setcounter{footnote}{0}
\baselineskip 0.58cm

\section{Introduction}

The high energy scale during inflation, presumably well beyond the reach
of the current and future particle accelerator experiments, calls for an
effective theory description of inflation
\cite{Cheung:2007st,Weinberg:2008hq}. This is because by construction
the effective field theory approach is systematic through which we can
account for our ignorance. A key observation in writing the effective
field theory of single-field inflation\footnote{
Extensions to the multi-field case are possible under certain constraints
\cite{multi-eft}.
} is to note that in the time-dependent
background the time translational symmetry is broken, while spatial
diffeomorphism is preserved \cite{Cheung:2007st}. The couplings that
determine the expansion of the effective theory of the Goldstone mode
$\pi$, which realizes the time diffeomorphism, are represented by a set
of mass scales $M_n^4$. In the so-called decoupling regime the Goldstone
$\pi$ could decouple from the metric fluctuations and the effective
action of $\pi$ is dramatically simplified. Especially, the first
expansion parameter $M_2^4$ is manifest in both quadratic and cubic
order of $\pi$: see \eqref{eq:Spi}.

The observation that the coefficient $M_2^4$ is common to the quadratic
and cubic action of $\pi$ indicates that, to leading order in the
decoupling limit, the corresponding correlation functions -- the power
spectrum and bispectrum -- are explicitly correlated. It means that
ideally, given an explicit analytic form of the power spectrum theoretically, we can
find unambiguously the corresponding bispectrum. Or, at the very least
observationally, it remains tantalizing because of the existence of
outliers in the power spectrum of the temperature fluctuations of the
cosmic microwave background~\cite{cmboutliers}. The explicit correlation would make
possible joint analysis using the two- and three-point correlation
functions~\cite{joint}, which can place much stronger constraints on cosmological
parameters. It can also open a compelling way of searching for new physics beyond
the paradigm of standard slow-roll inflation, since any deviations would strongly
signal the typical mass scale associated with new physics~\cite{newscale}.

In this article, we derive a direct and explicit relation between the
power spectrum and bispectrum of the primordial curvature perturbation
using the Goldstone mode $\pi$. Such a correlation was first explicitly
studied in the top-down approach \cite{Achucarro:2012fd} and
expanded into more general context in~\cite{Gong:2014spa}, in which heavy
degrees of freedom are integrated out to lead to an effective single
field description of inflation~\cite{cs-heavy} (see also~\cite{heavyeffects}).
To leading order of the heavy mass scale, the speed of sound $c_s$ uniquely
characterizes the effects of the heavy degrees of freedom
\cite{cs-heavy}, i.e. the coefficients of the effective
theory. Our approach here is conversely bottom-up, complementary to the
previous studies as we will see in the main text.

The article is organized as follows. In the next section, after briefly
reviewing the effective field theory of inflation, we derive the simple
expression of the correction to the power spectrum. By inverting it we
can write the unknown, model-dependent effective theory parameter in
terms of the power spectrum which can be constrained observationally.
In Section~\ref{sec:correlation}, we derive a direct and explicit
relation between the corrections of the power spectrum and bispectrum.
In Section~\ref{sec:squeezed} we discuss the consistency relation of the
squeezed bispectrum \cite{consistency} and the validity of the
correlation we derive. The final section is devoted to summary and
conclusions.

\section{Effective theory and correction to power spectrum}
\label{sec:P}

In this section, after briefly reviewing the effective field theory of inflation, 
we give the formula of the correction to the power spectrum due to
the deviation from usual slow-roll phase parametrized by the
expansion coefficient of the effective theory.

\subsection{Brief review of effective field theory of inflation}

We begin with a brief review of the effective field theory of
inflation~\cite{Cheung:2007st}. In unitary gauge, 
the information on the primordial curvature perturbation is encoded in
geometrical quantities respecting the time-dependent spatial
diffeomorphism symmetry. Then, the action for the primordial curvature
perturbation is written in general as
\begin{equation}
  S = \int d^4 x \sqrt{-g} F(g^{\mu\nu}, g_{\mu\nu}, 
  K_{\mu\nu}, R_{\mu\nu\rho\sigma}, \nabla_{\mu}, t) \, ,
\end{equation}
where $K_{\mu\nu}$ is the extrinsic curvature with respect to $t=$ constant
hypersurface. Since the zeroth and first order terms are determined by the
background quantities, the action can be expanded as
\begin{align}
S = \int d^4 x \sqrt{-g} & \left\{ \frac12 \mpl^2 R + \mpl^2 \dot{H}(t) g^{00} 
- \mpl^2 \left[ 3H^2(t)+\dot{H}(t) \right] 
\right. 
\nonumber\\         
& + F \left( \delta g^{00},\delta K_{\mu\nu},
             \delta R_{\mu\nu\rho\sigma} ; 
             g_{\mu\nu},g^{\mu\nu},\nabla_{\mu},t \right) \bigg\} \, ,
\end{align}
where $F$ represents second and higher order perturbation terms and
is given by
\begin{align}
F & = \frac{1}{2} M_2^4(t) \left( \delta g^{00} \right)^2 
                      + \frac{1}{3!} M_3^4(t) \left( \delta g^{00} \right)^3 + \cdots
\nonumber\\
& \quad    - \frac12 \bar{M}_1^3(t) \delta g^{00} \delta K
                      - \frac12 \bar{M}_2^3(t) K \delta K
                      - \frac12 \bar{M}_3^3(t) K^{\mu}{}_{\nu} \delta K^{\nu}{}_{\mu} 
                      - \cdots \, ,            
\end{align}
with $K \equiv K^\mu{}_\mu$.
It is noticed that time diffeomorphism invariance is broken in this
action. But, it can be recovered by the introduction of the
St\"uckelberg field $\pi(x)$, which corresponds to the Nambu-Goldstone boson
and transforms 
under the coordinate transformations  $t \rightarrow \widetilde{t} = t +\xi^0(x)$ and 
$\mathbi{x} \rightarrow \widetilde{\mathbi{x}}$ as
\begin{equation}
  \pi(x) \longrightarrow \widetilde{\pi}(\widetilde{x}(x)) 
  = \pi(x) - \xi^0(x) \, .
\end{equation}
In the decoupling regime $H \gtrsim M_2^2/\mpl$,
the action reduces to
\begin{equation}
\label{eq:Spi}
S_\pi = \int d^4x \sqrt{-g} \left\{ \frac{\mpl^2}{2}R - \mpl^2\dot{H} \left[ \dot\pi^2 - \frac{(\nabla\pi)^2}{a^2} \right] 
+ 2M_2^4 \left[ \dot\pi^2 + \dot\pi^3 - \dot\pi\frac{(\nabla\pi)^2}{a^2}
	 \right] - \frac{4}{3}M_3^4\dot\pi^3 + \cdots \right\} \, ,
\end{equation}
where the dots represent the higher derivative terms. The sound velocity
$c_s$ is related to $M_2$ as 
\begin{equation}
\label{eq:cs}
c_s^{-2} = 1 - \frac{2M_2^4}{\mpl^2\dot{H}} \, .
\end{equation}
In this article, we further set $M_3(t) = 0$ because $M_3^4 \sim
(1-c_s^{-2}) M_2^4$ on general arguments \cite{Senatore:2009gt}.  $\pi$
and $\calR$ are related to linear order by $\pi = -\calR/H$, so in the
regime \eqref{eq:Spi} is valid we can to first approximation consider
$\dot\pi \approx -\dot\calR/H$.

\subsection{Corrections to the power spectrum}
\label{subsec:power}

We first concentrate on the quadratic part and evaluate the correction
to power spectrum originating from the term with $M_2^4$. 
Since the standard slow-roll terms multiplied by $\dot{H}$ in \eqref{eq:Spi}
are dominant as various observations indicate, we treat
the quadratic contribution of $M_2^4$ as perturbation. In terms of the
speed of sound \eqref{eq:cs}, we assume that for a limited duration
$c_s$ deviates from unity, with the deviation being not too far away
from unity.  
Neglecting the metric perturbation as we consider the
decoupling regime so that $\sqrt{-g} = a^3$ simply, from \eqref{eq:Spi}
the quadratic part other than the usual slow-roll, which we may call second
order interaction, is
\begin{equation}
S_\text{2,int} = \int d^4x a^3 2M_2^4(t)\dot\pi^2 \, .
\label{eq:2int}
\end{equation}
The interaction Hamiltonian is then\footnote{One should be careful when the
interaction Lagrangian includes derivative terms. Conjugate momentum must
be defined by use of the full Lagrangian rather than the free part.}
\begin{equation}
\label{eq:Hint-2}
H_\text{int} = \int d^3x a^3 (-2) c_s^2 M_2^4(t)\dot\pi^2 
             \approx \int d^3x a^3 (-2) M_2^4(t)\dot\pi^2 \, ,
\end{equation}
where we have used the assumption that $c_s$ is not too far away from
unity. This interaction Hamiltonian can be expressed in terms of the
Fourier mode as
\begin{equation}
H_\text{int} = -2aM_2^4 \int \frac{d^3q_1d^3q_2}{(2\pi)^3} \delta^{(3)}(-\mathbi{q}_{12}) \pi_{\mathbi{q}_1}'\pi_{\mathbi{q}_2}' \, ,
\end{equation}
where $\mathbi{q}_{12\cdots n} \equiv \mathbi{q}_1+\mathbi{q}_2+\cdots+\mathbi{q}_n$, a prime
represents a derivative with respect to the conformal time $d\tau =
dt/a$, and
\begin{equation}
   \pi(\tau,\mathbi{x}) = \int \frac{d^3q}{(2\pi)^3} e^{i\mathbi{q}\cdot\mathbi{x}} \pi_{\mathbi{q}}(\tau). 
\end{equation}
Now we can compute the corrections using the standard in-in
formalism. We can straightly obtain
\begin{align}
\label{eq:Ppi-corr}
\Delta \left\langle \pi_{\mathbi{k}_1}\pi_{\mathbi{k}_2}(\tau) \right\rangle
& \equiv (2\pi)^3 \delta^{(3)}(\mathbi{k}_{12}) \frac{2\pi^2}{k_1^3} \Delta\calP_\pi
\nonumber\\
& = i \int_{\tau_0\to-\infty}^{\tau\to0} a d\tau' \left\langle 0 \left| \left[ H_\text{int}(\tau'), \pi_{\mathbi{k}_1}\pi_{\mathbi{k}_2}(\tau) \right] \right| 0 \right\rangle
\nonumber\\
& = (2\pi)^3 \delta^{(3)}(\mathbi{k}_{12})
2\Re \left[ 2i \widehat{\pi}_{{k}_1}^*\widehat{\pi}_{{k}_2}^*(0)
\int_{-\infty}^0 d\tau \left( -2a^2M_2^4 \right) \widehat{\pi}_{{k}_1}'\widehat{\pi}_{{k}_2}'(\tau) \right] \, ,
\end{align}
where we have expanded the free field $\pi_\mathbi{k}$ using the creation and annihilation operators as
\begin{equation}
\label{eq:Roperator}
\pi_\mathbi{k} = a_\mathbi{k} \widehat{\pi}_k + a_{-\mathbi{k}}^\dag \widehat{\pi}_k^* 
\quad \text{with} \quad
\left[ a_\mathbi{k}, a_\mathbi{q}^\dag \right] = (2\pi)^3 \delta^{(3)}(\mathbi{k}-\mathbi{q}) \, ,
\end{equation}
and $\widehat{\pi}_{k}(\tau)$ is the mode function solution given by
\begin{equation}
\label{eq:pi-sol}
\widehat{\pi}_k(\tau) = -\frac{\widehat{\calR}_k}{H} 
= \frac{-i}{\sqrt{4\epsilon k^3}\mpl} (1+ik\tau) e^{-ik\tau} \, .
\end{equation}
Thus, we immediately find the correction to the power spectrum as 
\begin{equation}
  \frac{\Delta\calP_\pi}{\calP_\pi}(k) \approx
 \frac{\Delta\calP_\calR}{\calP_\calR}(k)
 \approx \frac{k}{\mpl^2\epsilon H^2}
\int_{-\infty}^0 d\tau \left( -2M_2^4 \right) \sin(2k\tau) \, ,
\label{eq:correction_p}
\end{equation}
where 
\begin{equation}
\calP_\pi = \frac{\calP_\calR}{H^2} = \frac{1}{8\pi^2\mpl^2\epsilon}
\end{equation}
is the featureless flat spectrum.

\subsection{Inverting the power spectrum}

For future convenience, let us return to \eqref{eq:Ppi-corr} and write
it in an alternative form. The real part is obtained by
adding the complex conjugate:
\begin{align}
\label{eq:Ppi-corr3}
\frac{2\pi^2}{k_1^3}\Delta\calP_\pi & = 2\Re \left[ 2i \widehat{\pi}_{{k}_1}^*\widehat{\pi}_{{k}_2}^*(0)
\int_{-\infty}^0 d\tau \left( -2a^2M_2^4 \right) \widehat{\pi}_{{k}_1}'\widehat{\pi}_{{k}_2}'(\tau) \right] 
\nonumber\\
& = 2i \widehat{\pi}_{{k}_1}^*\widehat{\pi}_{{k}_2}^*(0)
\int_{-\infty}^0 d\tau \left( -2a^2M_2^4 \right) \widehat{\pi}_{{k}_1}'\widehat{\pi}_{{k}_2}'(\tau) + c.c. \, .
\end{align}
By noting from \eqref{eq:pi-sol} that
$\widehat{\pi}_k(-\tau) = -\widehat{\pi}_k^*(\tau)$ and $\widehat{\pi}_k'(-\tau) = {\widehat{\pi}_k}^*(\tau)$,
and by oddly extending $M_2^4$ to {\em define} $\widetilde{M}_2^4$ as
\begin{equation}
\widetilde{M}_2^4(\tau) \equiv \left\{
\begin{array}{ll}
M_2^4(\tau) & \text{ if } \tau < 0
\\
-M_2^4(-\tau) & \text{ if } \tau > 0
\end{array}
\right. \, ,
\end{equation}
\eqref{eq:Ppi-corr3} can be written as\footnote{
Notice that we are at this stage not directly computing the propagator by
adopting the $i\varepsilon$ prescription of the contour, which remains unchanged
though. Our goal is to invert \eqref{eq:correction_p} by incorporating mathematical manipulations in 
such a way that the model-dependent parameter $M_2^4$ is given in terms of 
$\Delta\calP_\pi$ which can be observationally constrained.
}
\begin{align}
\frac{2\pi^2}{k_1^3}\Delta\calP_\pi & = 2i \widehat{\pi}_{{k}_1}^*\widehat{\pi}_{{k}_2}^*(0)
\int_{-\infty}^\infty d\tau \left( -2a^2\widetilde{M}_2^4 \right) \widehat{\pi}_{{k}_1}'\widehat{\pi}_{{k}_2}'(\tau)
\nonumber\\
& = 2\pi^2\calP_\pi \frac{1}{k_1k_2} \frac{1}{2} 
\int_{-\infty}^\infty d\tau \frac{-2\widetilde{M}_2^4}{\epsilon\mpl^2H^2} ie^{-ik_{12}\tau} \, .
\end{align}
Since we have defined $\widetilde{M}_2^4$ oddly, only the odd part of $e^{-ik_{12}\tau}$ survives and finally we have, setting $k_1=k_2=k$,
\begin{equation}
\label{eq:Ppi-corr-final}
\frac{\Delta\calP_\pi}{\calP_\pi} = \frac{k}{2} 
\int_{-\infty}^\infty d\tau \frac{-2\widetilde{M}_2^4}{\epsilon\mpl^2H^2} \sin(2k\tau) \, .
\end{equation}

From \eqref{eq:Ppi-corr-final} we can write the coefficient $\widetilde{M}_2^4$, which is essentially $M_2^4$ in the effective action \eqref{eq:Spi}, 
in terms of $\Delta\calP_\pi/\calP_\pi$ as follows. From
$\sin(2k\tau) = \left( e^{2ik\tau} - e^{-2ik\tau} \right)/(2i)$,
we can multiply $e^{2ik\tau'}$ to both sides of \eqref{eq:Ppi-corr-final}
and integrate over $k$ to obtain
\begin{equation}
\int_{-\infty}^\infty dk e^{2ik\tau'} \frac{2i}{k} \epsilon\mpl^2H^2 \frac{\Delta\calP_\calR}{\calP_\calR}(k)
= \frac{1}{2} \int_{-\infty}^\infty d\tau \left( -2\widetilde{M}_2^4 \right)
\int_{-\infty}^\infty dk \left[ e^{2ik(\tau+\tau')} -
 e^{-2ik(\tau-\tau')} \right] 
= 2\pi \widetilde{M}_2^4(\tau') \, .
\end{equation}
Thus,
\begin{equation}
\label{eq:coeff-P}
2\widetilde{M}_2^4(\tau) = i\frac{2\epsilon\mpl^2H^2}{\pi} 
\int_{-\infty}^\infty \frac{dk}{k} \frac{\Delta\calP_\calR}{\calP_\calR}(k) e^{2ik\tau} \, .
\end{equation}
This is the inverse formula, in which $M_2^4$ can be expressed in terms
of the correction to power spectrum.

\section{Correlation between power spectrum and bispectrum}
\label{sec:correlation}

In this section, we first give the formula of bispectrum coming from
the cubic action \eqref{eq:Spi}, 
and then derive the explicit relation between
the correction to the power spectrum and the bispectrum.

\subsection{Bispectrum}

As advertised before, we only consider
the cubic order action with the coefficient $M_2^4$:
\begin{equation}
S_{3} = \int d^4x \sqrt{-g} 2M_2^4 \left[ \dot\pi^3 - \dot\pi\frac{(\nabla\pi)^2}{a^2} \right] \, .
\end{equation}
We can follow the same steps as before: the interaction Hamiltonian is
\begin{align}
H_\text{int} & = -\int d^3x a^3 \cdot 2M_2^4 \left[ \dot\pi^3 - \dot\pi\frac{(\nabla\pi)^2}{a^2} \right]
\nonumber\\
& = -2a^3M_2^4 \int \frac{d^3q_1d^3q_2d^3q_3}{(2\pi)^{3\cdot2}} \delta^{(3)}(-\mathbi{q}_{123})
\left[ \dot\pi_{\mathbi{q}_1}\dot\pi_{\mathbi{q}_2}\dot\pi_{\mathbi{q}_3}
+ \frac{\mathbi{q}_1\cdot\mathbi{q}_2}{3a^2} \pi_{\mathbi{q}_1}\pi_{\mathbi{q}_2}\dot\pi_{\mathbi{q}_3} + \text{2 perm} \right] \, .
\end{align}
Then, the bispectrum of $\pi$ becomes
\begin{align}
\label{eq:B-pi0}
\left\langle \pi_{\mathbi{k}_1}\pi_{\mathbi{k}_2}\pi_{\mathbi{k}_3}(\tau) \right\rangle
& \equiv (2\pi)^3 \delta^{(3)}(\mathbi{k}_{123}) B_\pi(\mathbi{k}_1,\mathbi{k}_2,\mathbi{k}_3)
\nonumber\\
& = (2\pi)^3 \delta^{(3)}(\mathbi{k}_{123}) \left\{
i \widehat{\pi}_{k_1}^*\widehat{\pi}_{k_2}^*\widehat{\pi}_{k_3}^*(0)
\int_{-\infty}^0 d\tau \left( -2aM_2^4 \right)  \right. 
\nonumber\\
& \qquad \qquad \qquad  \qquad \times
\Bigl[ 6\widehat{\pi}_{q_1}'\widehat{\pi}_{q_2}'\widehat{\pi}_{q_3}'(\tau) 
+ 2(\mathbi{k}_1\cdot\mathbi{k}_2)
\widehat{\pi}_{q_1}\widehat{\pi}_{q_2}\widehat{\pi}_{q_3}'(\tau) +
\text{2 perm} \Bigr] + c.c. \biggr\} \, .
\end{align}
Again, we can find that by extending $M_2^4$ oddly the complex conjugate includes the integral from 0 to $\infty$, so
\begin{equation}
\label{eq:B-pi}
B_\pi(\mathbi{k}_1,\mathbi{k}_2,\mathbi{k}_3) = 
i \widehat{\pi}_{k_1}^*\widehat{\pi}_{k_2}^*\widehat{\pi}_{k_3}^*(0) \int_{-\infty}^\infty d\tau \left( -2a\widetilde{M}_2^4 \right)
\left[ 6\widehat{\pi}_{q_1}'\widehat{\pi}_{q_2}'\widehat{\pi}_{q_3}'(\tau) 
+ 2(\mathbi{k}_1\cdot\mathbi{k}_2)
\widehat{\pi}_{q_1}\widehat{\pi}_{q_2}\widehat{\pi}_{q_3}'(\tau) +
\text{2 perm} \right] \, .
\end{equation}

\subsection{Bispectrum in terms of the power spectrum}

In this subsection, we can use \eqref{eq:coeff-P} and write the
bispectrum \eqref{eq:B-pi} purely in terms of the power spectrum and its
derivatives. Let us first consider the first term of \eqref{eq:B-pi}. We
can straightforwardly write, with $K \equiv k_{123}$,
\begin{align}
\label{eq:B-pi_term1}
i \widehat{\pi}_{k_1}^*\widehat{\pi}_{k_2}^*\widehat{\pi}_{k_3}^*(0) & \int_{-\infty}^\infty d\tau \left( -2a\widetilde{M}_2^4 \right)
6\widehat{\pi}_{q_1}'\widehat{\pi}_{q_2}'\widehat{\pi}_{q_3}'(\tau)
= (2\pi^2\calP_\pi)^2
\frac{H}{\pi} \frac{3H}{k_1k_2k_3} \int_{-\infty}^\infty \frac{dk}{k} \frac{\Delta\calP_\calR}{\calP_\calR}(k)
\int_{-\infty}^\infty d\tau \tau^2 e^{i(2k-K)\tau} 
\nonumber\\
& = (2\pi^2\calP_\pi)^2 \frac{3}{4}H \frac{1}{k_1k_2k_3} 
\int_{-\infty}^\infty \frac{dk}{k} \frac{\Delta\calP_\calR}{\calP_\calR}(k) \frac{d^2}{dk^2} \delta\left( k-\frac{K}{2} \right)
\nonumber\\
& = (2\pi^2\calP_\pi)^2 \frac{3}{4}H \frac{1}{k_1k_2k_3} 
\frac{d^2}{dk^2} \left. \left[ \frac{1}{k} \frac{\Delta\calP_\calR}{\calP_\calR}(k) \right] \right|_{k=K/2} \, ,
\end{align}
where for the second equality we have replaced $\tau^2$
in the time integral with two derivatives with respect to $k$,
and for the last equality we have iteratively integrated by
parts.

To proceed further, with $\calP_\calR^\text{(total)} = \calP_\calR + \Delta\calP_\calR$, from
\begin{equation}
\log\calP_\calR^\text{(total)} \approx \log\calP_\calR + \frac{\Delta\calP_\calR}{\calP_\calR} \, ,
\end{equation}
with $\calP_\calR$ being flat, we can find the spectral index and the
running respectively as\footnote{In case one takes into
account the slight tilt of $\calP_\calR$, the spectral index and the
running given here represent only the effect of $\Delta\calP_\calR/\calP_\calR$. Since we
assumed that for a limited duration $c_s$ deviates from unity, we can
separate the correction part from the standard slow-roll part, for both
of which, the spectral index and the running can be defined,
respectively.}
\begin{align}
n_\calR-1 & \equiv \frac{d\log\calP_\calR^\text{(total)}}{d\log{k}} = k\frac{d}{dk} \left( \frac{\Delta\calP_\calR}{\calP_\calR} \right) \, ,
\\
\alpha_\calR & \equiv \frac{dn_\calR}{d\log{k}} = k^2\frac{d^2}{dk^2} \left( \frac{\Delta\calP_\calR}{\calP_\calR} \right) 
+ k\frac{d}{dk} \left( \frac{\Delta\calP_\calR}{\calP_\calR} \right) \, .
\end{align}
Thus \eqref{eq:B-pi_term1} can be now written as
\begin{align}
& i \widehat{\pi}_{k_1}^*\widehat{\pi}_{k_2}^*\widehat{\pi}_{k_3}^*(0) \int_{-\infty}^\infty d\tau \left( -2a\widetilde{M}_2^4 \right)
6\widehat{\pi}_{q_1}'\widehat{\pi}_{q_2}'\widehat{\pi}_{q_3}'(\tau) 
\nonumber\\ 
& \qquad \qquad  = (2\pi)^4\calP_\pi^2 \frac{3}{2}H \frac{1}{k_1k_2k_3} \frac{1}{K^3}
\left[ \alpha_\calR - 3(n_\calR-1) + 2\frac{\Delta\calP_\calR}{\calP_\calR} \right] \bigg|_{k=K/2} \, .
\end{align}
We can proceed in a similar manner for the second term of \eqref{eq:B-pi} 
and find
\begin{align}
& i \widehat{\pi}_{k_1}^*\widehat{\pi}_{k_2}^*\widehat{\pi}_{k_3}^*(0) \int_{-\infty}^\infty d\tau \left( -2a\widetilde{M}_2^4 \right)
2(\mathbi{k}_1\cdot\mathbi{k}_2) \widehat{\pi}_{q_1}\widehat{\pi}_{q_2}\widehat{\pi}_{q_3}'(\tau)
\nonumber\\
& = (2\pi)^4\calP_\pi^2 \frac{1}{H} \frac{-\mathbi{k}_1\cdot\mathbi{k}_2}{(k_1k_2)^3k_3}
\left[ \left( 1 + \frac{k_{12}}{K} + \frac{2k_1k_2}{K^2} \right) \frac{\Delta\calP_\calR}{\calP_\calR} 
+ \left( -\frac{k_{12}}{K} - \frac{3k_1k_2}{K^2} \right) (n_\calR-1) +
 \frac{k_1k_2}{K^2} \alpha_\calR \right] \bigg|_{k=K/2} \, .
\end{align}

Thus, the bispectrum can be expressed in terms of the correction to
power spectrum, its first and second derivatives as
\begin{align}
\label{eq:bi-final}
& B_\pi(\mathbi{k}_1,\mathbi{k}_2,\mathbi{k}_3)  
\nonumber\\
& = (2\pi)^4\calP_\pi^2 \frac{H}{(k_1k_2k_3)^3}
\left[ {A}(k_1,k_2,k_3) \frac{\Delta\calP_\calR}{\calP_\calR} 
+ {B}(k_1,k_2,k_3) (n_\calR-1) 
+ {C}(k_1,k_2,k_3) \alpha_\calR \right] \bigg|_{k=K/2} \, ,
\end{align}
where the functions of momenta ${A}$, ${B}$ and ${C}$ are
given by, respectively,
\begin{align}
{A}(k_1,k_2,k_3) &= 
                  - \frac{1}{K^2}\sum_{i \ne j} k_i^2 k_j^3 
                  + 2 \frac{1}{K}\sum_{i > j} k_i^2 k_j^2
                  - \frac14 \sum_{i} k_i^3 \, , 
\\
{B}(k_1,k_2,k_3) &=
                   2 \frac{1}{K^2}\sum_{i \ne j} k_i^2 k_j^3 
                 - 3 \frac{1}{K}\sum_{i > j} k_i^2 k_j^2
                 + \frac14 \sum_{i \ne j} k_i k_j^2
                 - \frac14 k_1 k_2 k_3 \, , 
\\
{C}(k_1,k_2,k_3) &=
                  - \frac{1}{K^2}\sum_{i \ne j} k_i^2 k_j^3 
                  + \frac{1}{K}\sum_{i > j} k_i^2 k_j^2
                  - \frac14 k_1 k_2 k_3 \, .
\end{align}
This expression is one of the main results in this article.

\begin{figure}
 \begin{center}
  \includegraphics[width=0.9\textwidth]{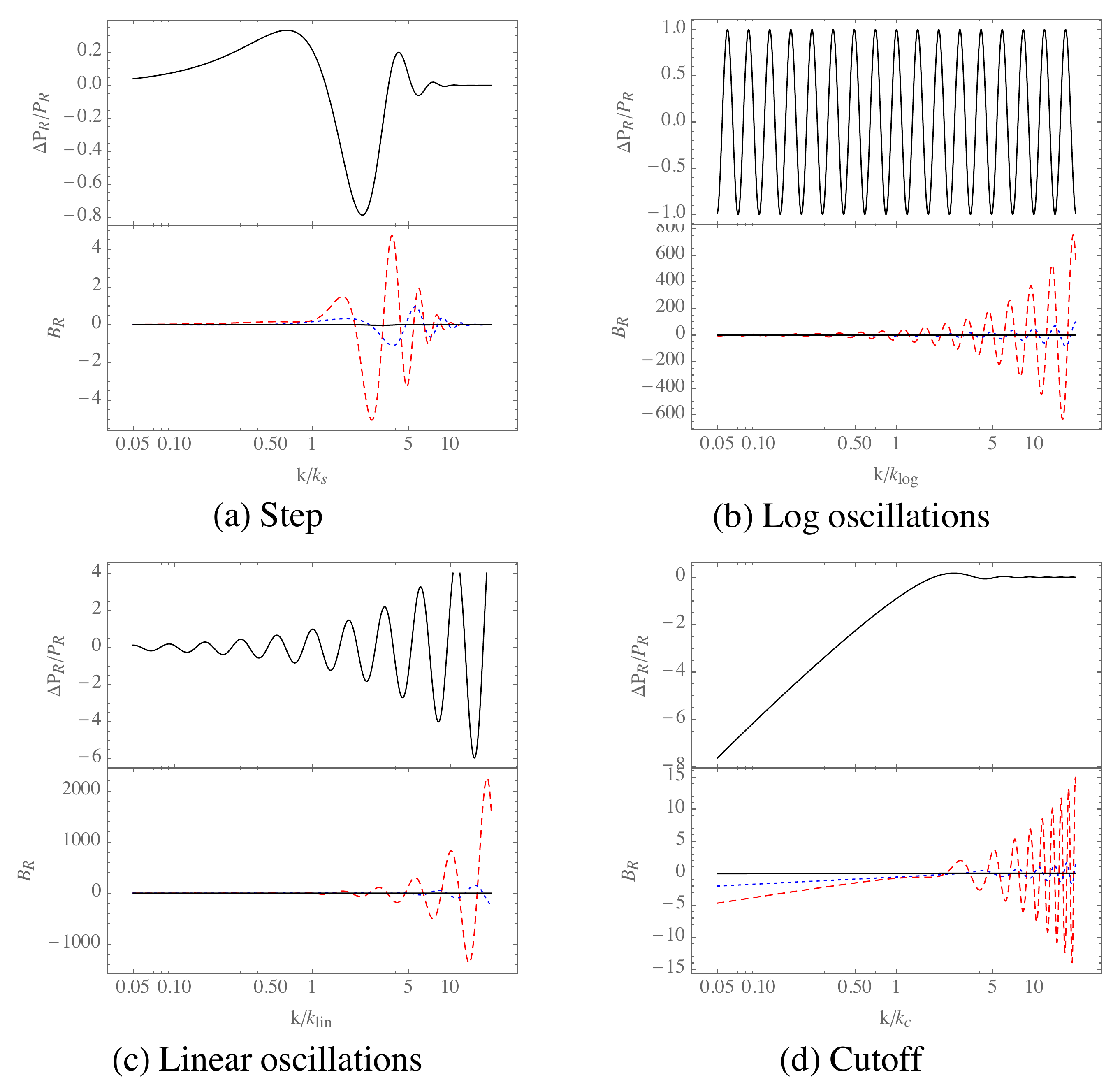}
 \end{center}
 \caption{(Upper panels) power spectrum and (lower panels) the corresponding 
 bispectrum for various feature models \eqref{eq:models} discussed in~\cite{Ade:2015lrj}. 
 For simplicity, we have set the amplitudes of the features as 
 $A_s = A_\text{log} = A_\text{lin} = 1$ and phases as $\varphi_\text{log} = \varphi_\text{in} = 0$. 
 For the step model, we have also set the damping scale $x_d = 1$. 
 Meanwhile, following~\cite{Ade:2015lrj} we have set $\log_{10}\omega_\text{log} = 1.25$, 
 $\log_{10}\omega_\text{lin} = 1.02$ and $n_\text{lin} = 0.66$. 
 We show the bispectrum projected onto the equilateral (red dashed), folded 
 (blue dotted) and squeezed (black solid) configurations.}
 \label{fig:corr}
\end{figure}

In Figure~\ref{fig:corr}, we show a few examples using the following
parametrized feature models~\cite{Ade:2015lrj}: a localized oscillatory burst
due to e.g. step in the inflaton potential, logarithmic and linear
oscillations and cutoff models given by
\begin{equation}
\label{eq:models}
\frac{\Delta\calP_\calR}{\calP_\calR} = \left\{
\begin{array}{ll}
A_s W_0\left(\dfrac{k}{k_s}\right) D\left(\dfrac{k/k_s}{x_d}\right)
& \text{(step)}
\\
A_\text{log} \cos\left[ \omega_\text{log}\log\left(\dfrac{k}{k_\text{log}}\right) + \varphi_\text{log} \right]
& \text{(logarithmic oscillations)}
\\
A_\text{lin} \left( \dfrac{k}{k_\text{lin}} \right)^{n_\text{lin}} \cos\left( \omega_\text{lin}\dfrac{k}{k_\text{lin}} + \varphi_\text{lin} \right)
& \text{(linear oscillations)}
\\
\log \left( \dfrac{\pi}{16} \dfrac{k}{k_c} |C_c-D_c|^2 \right)
& \text{(cutoff model)}
\end{array}
\right.
\, ,
\end{equation}
where the functions that appear in these parametrized feature models are
\begin{align}
W_0(x) & = \frac{1}{2x^4} \Big[ \left( 18x - 6x^3 \right) \cos(2x) + \left( 15x^2 - 9 \right) \sin(2x) \Big] \, ,
\\
D(x) & = \frac{x}{\sinh{x}} \, ,
\\
C_c & = \exp\left( \frac{-ik}{k_c} \right) \left[ H_0^{(2)}\left( \frac{k}{2k_c} \right) - \left( \frac{1}{k/k_c}+i \right) H_1^{(2)} \left( \frac{k}{2k_c} \right) \right] \, ,
\\
D_c & = \exp\left( \frac{ik}{k_c} \right) \left[ H_0^{(2)}\left( \frac{k}{2k_c} \right) - \left( \frac{1}{k/k_c}-i \right) H_1^{(2)} \left( \frac{k}{2k_c} \right) \right] \, ,
\end{align}
with $H_n^{(2)}$ being the Hankel function of the second kind.
As we can see, each power spectrum gives distinctively different patterns of 
the corresponding bispectrum in various configurations.

\section{Squeezed bispectrum and consistency relation}
\label{sec:squeezed}

We can note that \eqref{eq:bi-final} vanishes in the squeezed limit,
say, $k_1 \approx k_2$ and $k_3\to0$.\footnote{It was recently claimed
that for local observers, the squeezed limit vanishes in single-field
inflation
\cite{obs-squeezed}. But in
this article we do not take such effects into account and hence the
consistency relation should hold if we would calculate it adequately.}
This seems to contradict the consistency relation between the power
spectrum and the squeezed limit of the bispectrum
\cite{consistency},
\begin{equation}
B_\calR(k_1,k_2,k_3) \underset{k_3\to0}{\longrightarrow} (1-n_\calR)P_\calR(k_1)P_\calR(k_3) \, ,
\end{equation}
because as \eqref{eq:correction_p} shows the power spectrum is well away
from featureless flat one, so the corresponding spectral index is
non-trivial. Indeed, in \cite{Achucarro:2012fd}, the consistency relation
is recovered for features caused by non-trivial speed of sound.

Let us first return to the quadratic action for the {\em curvature perturbation}. Including the speed
of sound, it is written as
\begin{equation}
\label{eq:S2R}
S_2 = \int d^4x a^3\mpl^2\epsilon \left[ \frac{\dot\calR^2}{c_s^2} - \frac{(\nabla\calR)^2}{a^2} \right] \, ,
\end{equation}
so there are two possible sources of departure from the usual canonical
slow-roll~\cite{featuresource}: $\epsilon$ and $c_s$. Let us consider these two cases
separately. Our goal here is to see the form of the corrections to the
power spectrum for each case. But this seems unclear, since the form of
the interaction part of the quadratic action -- just $\dot\calR^2$ for
$c_s$, and $\dot\calR^2$ and $(\nabla\calR)^2$ for $\epsilon$ -- is different. 
Thus naively thinking the resulting correction terms would be of different structure.
We first assume that $c_s$ solely supplies the deviations from the
standard slow-roll in such a way that for a limited duration $c_s$
deviates from unity, with the deviation being not too far away from
unity. We may then write, with the canonical slow-roll part being the
leading, free part,
\begin{equation}
S_2 = \int d^4x a^3\mpl^2\epsilon \left[ \dot\calR^2 - \frac{(\nabla\calR)^2}{a^2} \right]
+ \underbrace{ \int d^4x a^3\mpl^2\epsilon \left( \frac{1}{c_s^2} - 1 \right) }_{\equiv
S_\text{2,int}} \, .
\end{equation}
Following the same steps as in Section~\ref{subsec:power}, we find
\begin{equation}
\label{eq:Pcorr-result1}
\frac{\Delta\calP_\calR}{\calP_\calR} = k \int_{-\infty}^0 d\tau \left( c_s^2-1 \right)
\sin(2k\tau) \, ,
\end{equation}
which is of the same structure as \eqref{eq:correction_p}.

For the case in which $\epsilon$ is responsible for the
departure from the standard slow-roll, let us split $\epsilon$
into the slowly varying part $\epsilon_0$ and the rapidly varying but
transient part $\Delta\epsilon$:
\begin{equation}
\epsilon = \epsilon_0 + \Delta\epsilon \, .
\end{equation}
We can rewrite $\Delta\epsilon$ as
\begin{equation}
\Delta\epsilon = \int \dot\epsilon dt 
 \approx H\epsilon_0 \int \eta dt
 \approx \epsilon_0\eta H \Delta t\, ,
\end{equation}
where $\Delta t = \calO(1/H)$ is the duration of departure and we have defined
another slow-roll parameter $\eta \equiv \dot\epsilon/(H\epsilon)$.
Then the quadratic action \eqref{eq:S2R}, with $c_s=1$ this time, can be written as
\begin{equation}
\label{eq:S2-epsilon-change}
S_2 = \int d^4x a^3\mpl^2\epsilon_0 \left[ \dot\calR^2 -
				     \frac{(\nabla\calR)^2}{a^2} \right]
+ \underbrace{ \int d^4x a^3\mpl^2 \Delta\epsilon \left[ \dot\calR^2 -
				       \frac{(\nabla\calR)^2}{a^2}
				      \right] }_{\equiv
S_\text{2,int}} \, ,
\end{equation}
and the corresponding correction to the power spectrum is
\begin{equation}
\label{eq:Pcorr-result2}
\frac{\Delta\calP_\calR}{\calP_\calR} = k \int_{-\infty}^0 d\tau
\left(-2 \frac{\Delta\epsilon}{\epsilon_0} \right) \sin(2k\tau) \, .
\end{equation}
Comparing this with \eqref{eq:Pcorr-result1}, we see
that two sources of the departure from the standard slow-roll
leads to the same structure of the correction as \eqref{eq:correction_p}.
This seems to suggest that indeed $M_2^4$ captures the deviation
from usual slow-roll on general ground.

We now return to our starting equation \eqref{eq:Spi} to clarify this
inconsistency. A key observation is that unlike $\calR$, which is frozen
on super-horizon scales, $\pi$ evolves as
\begin{equation}
\label{eq:pi-R}
\dot\pi = -\frac{\dot\calR}{H} - \epsilon\calR + \frac{\dot\calR^2}{H^2} 
+ 3\epsilon \frac{\dot\calR\calR}{H} + \cdots \, ,
\end{equation}
where the non-linear terms follow from the fact that essentially
$\pi$ is the time translation between spatially flat and comoving
hypersurfaces~\cite{Noh:2004bc}. Also we have omitted terms that are further
suppressed in slow-roll parameters. Taking into account the 
sub-leading terms in $\dot\pi$, at quadratic order of the curvature
perturbation $M_2^4$ contributes\footnote{
The standard slow-roll terms, multiplied by $\dot{H}$ in \eqref{eq:Spi},
also give rise to additional sub-leading terms, but they are $\calO(\epsilon^2)$
so we do not include them here.
}
\begin{equation}
S_\calR \supset \int d^4 a^3 \left[ \frac{2M_2^4}{H^2}\dot\calR^2
+ 2\epsilon \left( -3M_2^4 + \frac{\dot{M}_2^4}{H} \right) \calR^2 \right] \, .
\end{equation}
Thus, the speed of sound of the curvature perturbation is identical to
that of $\pi$ given by \eqref{eq:cs}. At the same time there do exist
changes in $\calR^2$ terms as \eqref{eq:S2-epsilon-change}, but they are
slow-roll suppressed. Since we have only considered the leading effects
that only capture the speed of sound, the bispectrum \eqref{eq:bi-final}
is enhanced in the equilateral configuration while it is not in the
squeezed limit. Indeed, by considering the sub-leading terms in
\eqref{eq:pi-R}, we have at cubic order new terms
$\dot\calR^2\calR$ and {$\calR(\nabla\calR)^2$} that lead to
non-vanishing bispectrum in the squeezed
limit~\cite{Cheung:2007sv}. More specifically, the new terms
{$\dot\calR^2\calR$} and {$\calR(\nabla\calR)^2$} in the cubic
order action {give} up to numerical coefficient
\begin{equation}
\frac{k_1k_2k_3}{k_1^3+k_2^3+k_3^3} \frac{(k_1k_2k_3)^2B_\calR(k_1,k_2,k_3)}{(2\pi)^4\calP_\calR^2}
\underset{k_3\to0}{\longrightarrow} \epsilon \frac{\Delta\calP_\calR}{\calP_\calR} \, ,
\end{equation}
with $k_1 \approx k_2 \equiv k$. Still the consistency relation is not
recovered, but this is because we are not taking into account all the
next-to-leading terms in the decoupling limit, {such as the
modification of the mode functions: terms of $\calO(1/c_s^2)$ and
$\calO(\epsilon/c_s^2)$ do not contribute to the squeezed limit while
only terms of $\calO(\epsilon)$ do}~\cite{RenauxPetel:2010ty}. Our
calculation is done only up to the leading order.

\section{Summary}

In this article, we have derived the direct relation between the
corrections of power spectrum and bispectrum of the primordial curvature
perturbation.  Our formula is based on the effective field theory
approach to inflation, which to first approximation captures the effects
of the non-trivial speed of sound. If we would observationally detect
the deviation from the standard slow-roll inflation, it is important to
check the relation derived here, which could prove/disprove whether such
a deviation can be attributed to the variation of sound velocity.

We have also shown that the corrections to the power spectrum from
non-trivial features of sound velocity and expansion rate of the
universe, which characterize the deviation from the standard slow-roll
inflation, have the same form. It is interesting to check whether we can
extend this kind of unified treatment to higher order correlation
functions. We have also discussed the squeezed limit of the bispectrum
and the consistency relation. In the leading order calculations we have
adopted in this article, the squeezed limit vanishes. But, if we take
into account sub-leading orders adequately, the consistency relation
would be recovered.

The next step is to include the sub-leading order effects such as the
terms beyond decoupling limit and the $M_3$ terms. Then, we will have
further (consistency) relation, which is useful to identify new physics
causing such a deviation.

\subsection*{Acknowledgments}

JG thanks Tokyo Institute of Technology for hospitality during ``Workshop on Particle Physics, Cosmology, and Gravitation'' where this work was initiated.
JG acknowledges the support from the Korea Ministry of Education, Science and Technology, 
Gyeongsangbuk-Do and Pohang City for Independent Junior Research Groups at the Asia Pacific Center for Theoretical Physics.
JG is also supported in part by a TJ Park Science Fellowship of POSCO TJ Park Foundation 
and the Basic Science Research Program through the National Research Foundation of
Korea (NRF) Research Grant NRF-2016R1D1A1B03930408. 
MY is supported in part by Japan Society for the Promotion of Science (JSPS) 
Grant-in-Aid for Scientific Research Nos. 25287054 and 26610062,
and the Grant-in-Aid for Scientific Research on Innovative Areas 
``Cosmic Acceleration'' No. 15H05888.

  
\end{document}